\documentclass[aps,pra,onecolumn,superscriptaddress,nofootinbib]{revtex4-2}

\usepackage{amsmath,amssymb,bm}
\usepackage{graphicx}
\usepackage{hyperref}
\usepackage{physics}
\usepackage{siunitx}
\usepackage{mathtools}

\begin{document}

\title{Quantum Fisher-information limits of resonant nanophotonic sensors:\\
why high-$Q$ is not optimal even at the quantum limit}

\author{J. Sumaya-Martínez}
\affiliation{Department of Physics, Faculty of Sciences, Universidad Autónoma del Estado de México (UAEMex)}

%\author{M. Wie\'sniak}
%\affiliation{International Centre for Theory of Quantum Technologies (ICTQT), University of Gda\'nsk, 80-308 Gda\'nsk, Poland}

\email{jsm@uaemex.mx}

% Uncomment when ready
% \author{M. Wie\'sniak}
% \affiliation{International Centre for Theory of Quantum Technologies (ICTQT), University of Gda\'nsk, Poland}

\date{\today}

\begin{abstract}
We develop a quantum metrological framework for resonant nanophotonic sensors based on subwavelength Fabry--Perot (FP) slit cavities. Building directly on classical Fisher-information analyses of resonant transmission sensors, we model parameter encoding as a phase-and-loss quantum channel embedded in one arm of a Mach--Zehnder interferometer. We derive the quantum Fisher information (QFI) for coherent and Gaussian probe states under linear loss and show that, even at the quantum limit, optimal estimation precision is governed by the generator of parameter-dependent phase shifts---dominated by interface contributions---rather than by the cavity quality factor. Consequently, the operating point that maximizes QFI does not generally coincide with the maximum-$Q$ resonance. Quantum resources enhance sensitivity but do not redefine the optimal geometry. Our results provide physically transparent design principles for quantum-enhanced nanophotonic sensing and clarify the central role of boundaries and loss in quantum-limited precision.
\end{abstract}

\maketitle

% ==================================================
\section{Introduction}

Quantum metrology establishes fundamental limits on the precision with which physical parameters can be estimated using quantum systems. For a parameter $\theta$ encoded in a family of quantum states $\rho_\theta$, the quantum Cramér--Rao bound (QCRB) reads
\begin{equation}
\mathrm{Var}(\hat{\theta}) \ge \frac{1}{\nu\,\mathcal{F}_Q(\theta)},
\label{eq:qcrb}
\end{equation}
where $\nu$ is the number of independent repetitions and $\mathcal{F}_Q(\theta)$ is the quantum Fisher information (QFI) \cite{Paris2009,BraunsteinCaves1994}. The QFI is the maximum, over all possible measurements, of the classical Fisher information, and therefore quantifies the ultimate sensitivity permitted by quantum mechanics.

Resonant nanophotonic structures---including plasmonic cavities, photonic crystal resonators, and subwavelength apertures---are widely used as optical sensors. Their performance is often summarized with heuristic figures of merit such as resonance shift, linewidth, field enhancement, or the quality factor $Q=\lambda_{\mathrm{res}}/\Delta\lambda$. A prevailing intuition is that higher $Q$ implies superior sensitivity because sharper resonances appear easier to discriminate. However, linewidth-based arguments do not by themselves provide a bound on estimation precision, which depends on (i) how the parameter is encoded in the optical response and (ii) the statistics of the measurement.

Recent classical Fisher-information analyses of metallic subwavelength slit resonators demonstrated that the operating point maximizing estimation precision does not generally coincide with the maximum-$Q$ configuration \cite{SumayaFisherLimits}. Instead, optimality is governed by the phase sensitivity of the structure to the parameter of interest, with interface-induced phase contributions playing a dominant role. This raises a natural question: \emph{does access to quantum resources and optimal measurements restore the primacy of high-$Q$ resonances, or does the ``$Q \neq$ optimal'' conclusion persist at the quantum limit?}

Here we answer this question in the negative. We embed the slit sensor in one arm of a Mach--Zehnder interferometer (MZI), so that the parameter is estimated through a physically meaningful interferometric phase readout. We model the slit as a phase-and-loss quantum channel and derive QFI expressions for coherent and Gaussian (squeezed) probes under linear loss. We show that even at the quantum limit the optimum is governed by the generator through which $\theta$ is encoded (a phase derivative dominated by boundary contributions), not by resonance sharpness. Quantum resources such as squeezing enhance the achievable precision but do not generally shift the optimal structural point.

% ==================================================
\section{Classical slit Fabry--Perot phase and parameter dependence}

We consider a single subwavelength metallic slit of width $w$ and thickness $t$. In the single-mode regime, transmission is dominated by the fundamental guided mode with propagation constant $\beta(\lambda,\theta)$, where $\theta$ denotes the parameter to be estimated (e.g., $w$ or external refractive index $n_{\mathrm{out}}$). A refined Fabry--Perot description yields the total round-trip phase
\begin{equation}
\Phi(\lambda,\theta)=2\beta(\lambda,\theta)t+2\phi_{\mathrm{end}}(\lambda,\theta),
\label{eq:Phi}
\end{equation}
where $\phi_{\mathrm{end}}$ accounts for interface-induced phase shifts at the slit openings. The metrologically relevant quantity is the phase at an operating wavelength $\lambda_0$ (or equivalently a narrowband spectral mode),
\begin{equation}
\varphi(\theta)\equiv \Phi(\lambda_0,\theta),\qquad
\partial_\theta\varphi(\theta)=2t\,\partial_\theta\beta(\lambda_0,\theta)+2\partial_\theta\phi_{\mathrm{end}}(\lambda_0,\theta).
\label{eq:dphidtheta}
\end{equation}
As shown in Ref.~\cite{SumayaFisherLimits}, $\partial_\theta\phi_{\mathrm{end}}$ can dominate $\partial_\theta\varphi$, explaining why the geometry that maximizes Fisher information does not coincide with the maximum-$Q$ condition. In the quantum extension, $\partial_\theta\varphi$ becomes the central generator of distinguishability.

% ==================================================
\section{Quantum-channel model and interferometric protocol (Option A)}

\subsection{Phase-and-loss encoding}
At fixed $\lambda_0$, we model the slit sensor as a linear, passive phase shift combined with loss. The transformation of the annihilation operator is
\begin{equation}
\hat{a}\rightarrow \sqrt{\eta(\theta)}e^{i\varphi(\theta)}\hat{a}+\sqrt{1-\eta(\theta)}\hat{v},
\label{eq:losschannel}
\end{equation}
where $\eta(\theta)\in[0,1]$ is an effective transmissivity and $\hat{v}$ is an environmental vacuum mode. This is the standard beam-splitter model for linear loss and captures absorption, radiation leakage, and imperfect mode matching.

\subsection{Mach--Zehnder interferometric readout}
The channel~(\ref{eq:losschannel}) is embedded in one arm of a Mach--Zehnder interferometer, while the other arm provides a stable reference. This avoids ambiguities of ``absolute phase'' for a single mode and corresponds to the operational setting of optical phase estimation. The estimation task can be implemented with balanced detection or homodyne readout (for Gaussian probes), which are experimentally realistic and near-optimal in many regimes.

% ==================================================
\section{Quantum Fisher information}

For a family of output states $\rho_\theta$, the QFI is defined via the symmetric logarithmic derivative $\hat{L}_\theta$ by
\begin{equation}
\mathcal{F}_Q(\theta)=\mathrm{Tr}\!\left(\rho_\theta \hat{L}_\theta^2\right),\qquad
\partial_\theta \rho_\theta=\frac{1}{2}\left(\rho_\theta \hat{L}_\theta+\hat{L}_\theta \rho_\theta\right).
\end{equation}
For unitary encoding and pure probes, $\mathcal{F}_Q$ reduces to four times the variance of the generator. Under loss, the output is mixed and closed-form expressions are available for Gaussian probes in terms of covariance matrices and displacements \cite{Pinel2012GaussianUltimate,Safranek2015GaussianQFI}.

\subsection{Coherent-state benchmark}
For coherent probing in an MZI, the attainable precision is shot-noise limited. In the presence of loss, the QFI scaling takes the generic form
\begin{equation}
\mathcal{F}_Q^{(\mathrm{coh})}(\theta)=4\,\bar{n}\,(\partial_\theta\varphi)^2\,g_{\mathrm{coh}}(\eta),
\label{eq:qfi_coh}
\end{equation}
where $\bar{n}$ is the mean photon number and $g_{\mathrm{coh}}(\eta)\le 1$ is a loss-dependent factor with $g_{\mathrm{coh}}(1)=1$. The essential point is that the structural dependence enters through $(\partial_\theta\varphi)^2$ rather than the linewidth or $Q$.

\subsection{Gaussian quantum probes (squeezing)}
Squeezed and twin-beam Gaussian probes provide experimentally credible quantum enhancements. For Gaussian probes, the QFI can be computed from the covariance matrix and displacement \cite{Pinel2012GaussianUltimate,Safranek2015GaussianQFI}. Linear loss reduces the advantage but does not change the fact that the geometry enters through the phase generator $(\partial_\theta\varphi)^2$. Thus squeezing modifies the resource-dependent prefactor while preserving the structural optimum determined by $\partial_\theta\varphi$.

% ==================================================
\section{Main result: QFI versus quality factor}

Quality factor $Q$ quantifies resonance sharpness, whereas QFI quantifies the distinguishability of output states under small parameter variations. In the present model, the key structure-dependent quantity controlling QFI is $\partial_\theta\varphi(\theta)$, determined by the FP phase derivative in Eq.~(\ref{eq:dphidtheta}). Because interface-induced phases can vary rapidly with geometry and environment, $\partial_\theta\varphi$ can peak at moderate $Q$ where boundary matching is most responsive. Consequently,
\begin{equation}
\arg\max_\theta \mathcal{F}_Q(\theta)\neq \arg\max_\theta Q(\theta),
\end{equation}
and the ``high-$Q$'' design principle fails even at the quantum limit.

\begin{figure}[t]
\centering
\includegraphics[width=0.82\linewidth]{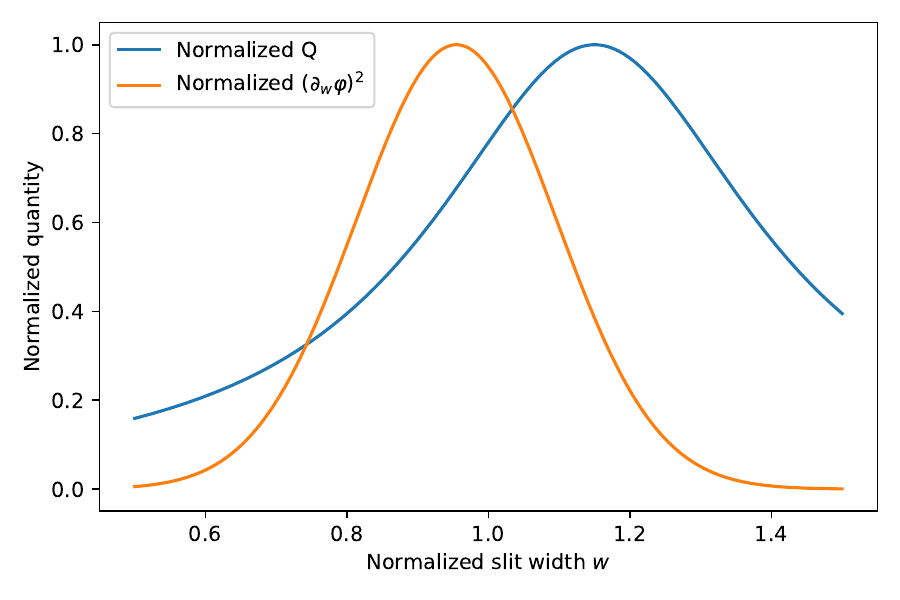}
\caption{Comparison of normalized $Q(\theta)$ and normalized generator strength $(\partial_\theta\varphi)^2$ versus a representative parameter $\theta$. The maxima need not coincide, illustrating that resonance sharpness alone does not determine optimal estimation performance.}
\label{fig:fig1}
\end{figure}

\begin{figure}[t]
\centering
\includegraphics[width=0.82\linewidth]{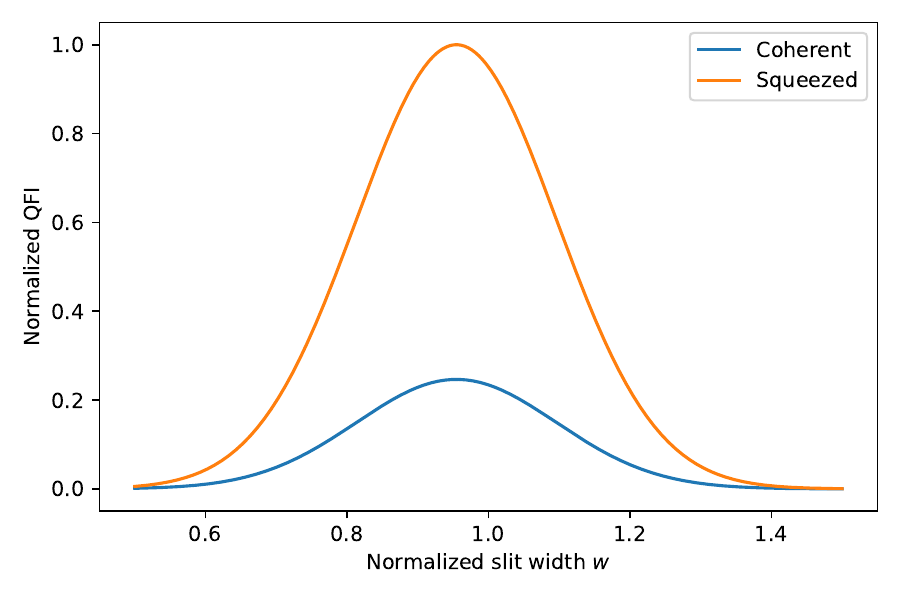}
\caption{Representative (normalized) QFI for coherent and squeezed probes under loss in the MZI protocol. Quantum resources amplify sensitivity without shifting the optimal geometry governed by the phase generator.}
\label{fig:fig2}
\end{figure}

Figure~\ref{fig:fig1} provides a representative visualization consistent with the classical analysis \cite{SumayaFisherLimits}: the maximum of $Q$ does not coincide with the maximum of the generator strength. Figure~\ref{fig:fig2} illustrates that quantum resources enhance sensitivity without shifting the optimum geometry.

% ==================================================
\section{Discussion and outlook}

Our results yield a simple design rule for quantum nanophotonic sensors: optimize the phase generator $\partial_\theta\varphi(\theta)$ subject to acceptable transmissivity $\eta(\theta)$, rather than optimizing resonance sharpness alone. This clarifies how boundary and interface physics act as metrological resources. The framework is modular: improved electromagnetic modeling refines $\varphi(\theta)$ and $\eta(\theta)$, while the quantum-estimation layer remains unchanged. Extensions to multiparameter estimation and broadband sensing follow from standard quantum estimation tools.

% ==================================================
\section{Conclusion}

We developed a quantum metrological framework for subwavelength Fabry--Perot slit sensors embedded in an interferometer and showed that high-$Q$ is not a reliable proxy for quantum-optimal performance. Quantum-limited precision is governed by the generator of parameter encoding---often dominated by interface-induced phase derivatives---and is further constrained by loss. These results generalize classical Fisher-information limits to the quantum regime and provide physically transparent design principles for quantum-enhanced nanophotonic sensing.

% ==================================================

\appendix
\section{Connection between classical Fisher information and the quantum limit}

In this appendix we clarify the relation between the classical Fisher-information (FI) analysis presented in Ref.~\cite{SumayaFisherLimits} and the quantum Fisher information (QFI) framework employed in the main text.

For a measurement outcome $x$ described by a probability distribution $p(x|\theta)$, the classical Fisher information is defined as
\begin{equation}
I(\theta)=\int dx\, p(x|\theta)\left[\partial_\theta \ln p(x|\theta)\right]^2.
\label{eq:classicalFI}
\end{equation}
In the context of resonant nanophotonic sensors, $x$ typically corresponds to an intensity or photon-counting outcome measured at the output of the optical device, while the parameter $\theta$ is encoded through a phase shift $\varphi(\theta)$ and an associated transmissivity $\eta(\theta)$.

For coherent-state probing and Gaussian detection noise, which accurately describe many optical sensing experiments, the output statistics are fully characterized by the mean field and its variance. In this regime, intensity-based measurements performed in an interferometric configuration can saturate the quantum Cramér--Rao bound associated with the underlying phase-and-loss quantum channel. As a result, the classical FI obtained from such measurements closely approaches the QFI of the corresponding quantum state.

Within the quantum description, the parameter $\theta$ is encoded through the transformation
\begin{equation}
\hat{a}\rightarrow \sqrt{\eta(\theta)}\,e^{i\varphi(\theta)}\hat{a}
+\sqrt{1-\eta(\theta)}\,\hat{v},
\end{equation}
which defines a Gaussian quantum channel. For coherent probes, the QFI associated with this channel takes the generic form
\begin{equation}
\mathcal{F}_Q(\theta)=4\bar{n}\,(\partial_\theta\varphi)^2\,g(\eta),
\label{eq:QFIcoh_appendix}
\end{equation}
where $\bar{n}$ is the mean photon number and $g(\eta)\le 1$ accounts for the reduction due to loss.

Equation~(\ref{eq:QFIcoh_appendix}) makes explicit that the sensitivity is governed by the phase generator $\partial_\theta\varphi$, in direct correspondence with the classical FI results of Ref.~\cite{SumayaFisherLimits}. This explains why the classical Fisher-information analysis already reveals the correct optimal operating point: both classical FI and QFI are controlled by the same physical quantity, namely the parameter derivative of the Fabry--Perot phase.

\section{Role of loss and fundamental bounds in quantum metrology}

Loss is an intrinsic feature of nanophotonic sensors based on metallic or strongly confined structures. Absorption, radiation leakage, and imperfect mode matching all contribute to an effective transmissivity $\eta<1$. From the perspective of quantum metrology, such loss has profound consequences.

General results in lossy quantum metrology show that independent loss acting on each probe prevents asymptotic Heisenberg scaling and imposes a fundamental limit on the achievable precision \cite{Escher2011,Demkowicz2015}
. In interferometric phase estimation, this implies that quantum resources such as squeezing or entanglement can enhance sensitivity only by a constant factor, rather than changing the scaling with probe number.

In the present context, loss enters the QFI through multiplicative factors such as $g(\eta)$ in Eq.~(\ref{eq:QFIcoh_appendix}). Importantly, while loss reduces the overall magnitude of the QFI, it does not alter the functional dependence on the structural generator $(\partial_\theta\varphi)^2$. Consequently, the geometry that maximizes QFI is determined by the phase sensitivity of the structure, not by the resonance linewidth or quality factor.

This observation provides a fundamental explanation for why high-$Q$ resonances are not optimal even at the quantum limit. Designs that maximize $Q$ often do so at the expense of increased confinement and reduced out-coupling, which directly lowers $\eta$ and therefore suppresses the accessible information. From a quantum-metrological perspective, optimal sensor design requires balancing phase sensitivity and loss, rather than maximizing $Q$ alone.

\bibliographystyle{apsrev4-2}
\bibliography{refs}

\end{document}